\documentclass[%
reprint,
superscriptaddress,
frontmatterverbose, 
amsmath,amssymb,aps,
]{revtex4-2}
\usepackage{xcolor}
\usepackage{graphicx}
\usepackage{dcolumn}
\usepackage{bm}
\usepackage{blindtext}
\usepackage{lipsum}
\usepackage{hyperref}
\usepackage[mathlines]{lineno}
\usepackage{amsmath}
\usepackage{multirow}
\usepackage{graphicx}
\usepackage{tabularx}
\usepackage{setspace}
\usepackage{makecell}
\usepackage{natbib}
\usepackage{tabularx}

\usepackage[
scale=0.85, marginratio={1:1, 2:3}, ignore all,
margin=0.65in,
]{geometry}

\begin{document}

\preprint{APS/123-QED}

\title{Hexagonal to Monoclinic Phase Transition in Dense Hydrogen Phase III Detected by High-Pressure NMR}

  \author{Meng Yang}
   \affiliation{Center for High-Pressure Science and Technology Advance Research, Beijing, China}

\author{Yishan Zhou}
   \affiliation{Center for High-Pressure Science and Technology Advance Research, Beijing, China}

\author{Rajesh Jana}
    \affiliation{Center for High-Pressure Science and Technology Advance Research, Beijing, China}

\author{Takeshi Nakagawa}
   \affiliation{Center for High-Pressure Science and Technology Advance Research, Beijing, China}

\author{Yunhua Fu}
   \affiliation{Center for High-Pressure Science and Technology Advance Research, Beijing, China}
   \affiliation{School of Earth and Space Sciences, Peking University, Beijing, China}
   
\author{Thomas Meier}
\email{thomas.meier@sharps.ac.cn}
\affiliation{Shanghai Key Laboratory MFree, Institute for Shanghai Advanced Research in Physical Sciences, Pudong, Shanghai, 201203}
\affiliation{Center for High-Pressure Science and Technology Advance Research, Beijing, China}

\date{\today}

\begin{abstract}
Conclusive crystal structure determination of the high pressure phases of hydrogen remains elusive due to lack of core electrons and vanishing wave vectors, rendering standard high-pressure experimental methods moot. \textit{Ab-initio} DFT calculations have shown that structural polymorphism might be solely resolvable using high-resolution nuclear magnetic resonance (NMR) spectroscopy at mega-bar pressures, however technical challenges have precluded such experiments thus far. Here, we present \textit{in-situ} high-pressure high-resolution NMR experiments in hydrogen phase III between 181 GPa and 208 GPa at room temperature. Our spectra suggest that at lower pressures phase III adopts a hexagonal $P6_1 22$ crystal structure, transitioning into a monoclinic \textit{$C2/c$} phase at about 197 GPa. The high resolution spectra are in excellent agreement with earlier structural and spectral predictions and underline the possibility of a subtle $P6_1 22 \rightarrow C2/c$ phase transition in hydrogen phase III. These experiments show the importance of a combination of \textit{ab-initio} calculations and low-Z sensitive spectral probes in high-pressure science in elucidating the structural complexity of the most abundant element in our universe.
\end{abstract}
\maketitle

Hydrogen exhibits a rich phase diagram under high pressures, making it a subject of intense study in condensed matter physics\cite{Geng2017,Gregoryanz2020}. Among the various phases, hydrogen phase III, observed at pressures above 150 GPa, has garnered significant attention due to its intriguing properties and potential implications for understanding fundamental physical phenomena and material science applications.
\\
The study of hydrogen under extreme conditions provides critical insights into the behavior of materials at high pressures and temperatures, relevant to both planetary science\cite{Mao2020,Hu2021a} and the quest for room-temperature superconductivity\cite{Mao2016, Lv2020, Mao2021}. Theoretical predictions and experimental discoveries have revealed that hydrogen, typically a diatomic molecular gas under ambient conditions, undergoes a series of phase transitions leading to complex crystal structures as pressure increases, influenced by an intricate interplay between electronic, vibrational, and structural properties of hydrogen molecules \cite{Silvera1980,Freiman2017a, Ackland2015, Gregoryanz2020}.
\\
Phase III of hydrogen is particularly notable for its complex and elusive crystal structure. Initial identification of this phase was based on anomalies in the Raman spectra and changes in the optical properties of hydrogen under pressures exceeding 150 GPa\cite{Howie2012, Howie2015}. Complimentary synchrotron X-ray diffraction studies have provided more detailed insights into the nature of phase III\cite{Ji2019}, suggesting the presence of a layered hexagonal structure, though the exact atomic arrangement remains ambiguous. Theoretical models have proposed several candidate structures for hydrogen phase III, ranging from hexagonal to monoclinic space groups, each with distinct molecular orientations and bonding characteristics\cite{Pickard2007, Monserrat2016}.
\\
However, conclusive crystal structure determination of phase III has been intangible due to the weak scattering cross-section of hydrogen atoms, making them almost invisible even under strong intense synchrotron radiation\cite{Stojilovic2012}. Moreover vanishing wave vectors and increasing luminescence of diamond anvils at these pressures make Raman or Infra-red spectroscopy challenging. 
\\
\textit{Ab-initio} density functional theory (DFT) calculations by Monserrat et al.\cite{Monserrat2019} pointed out that NMR spectroscopy could be the sole distinct spectroscopic method with which high pressure polymorphs of hydrogen might be experimentally resolvable, due to its ability to sense minuscule local magnetic field fluctuations and its high sensitivity to hydrogen atoms\cite{Levitt2000, Slichter2013}.\\
The two most-likely possible crystal structure candidates for phase III are a hexagonal $P6_1 22$ phase and a monoclinic $C2/c$ atomic arrangement with 36 and 24 atoms per unit cell respectively. The former exhibits six distinct crystallographically non-equivalent sites occupied by six hydrogen atoms each, leading to four different chemical shielding values, with two chemical shieldings merging. Thus a triplet structure emerges in the $^1H$-NMR spectra with a pronounced intensity ratio of $2:3:1$ (from low to high values of chemical shielding). The monoclinic structural candidate, $C2/c$, also exhibits six distinct sites with four atoms per site, and a degeneracy of chemical shieldings into two pairs of 12 atoms each, yielding a symmetric doublet structure in the $^1H$-NMR frequency range. Thus, both structures are reasonably distinct from each other and could be used to conclusively identify the high pressure structure of phase III.
\\
\begin{figure*}[ht]
\centering    
\includegraphics[width=.75\textwidth]{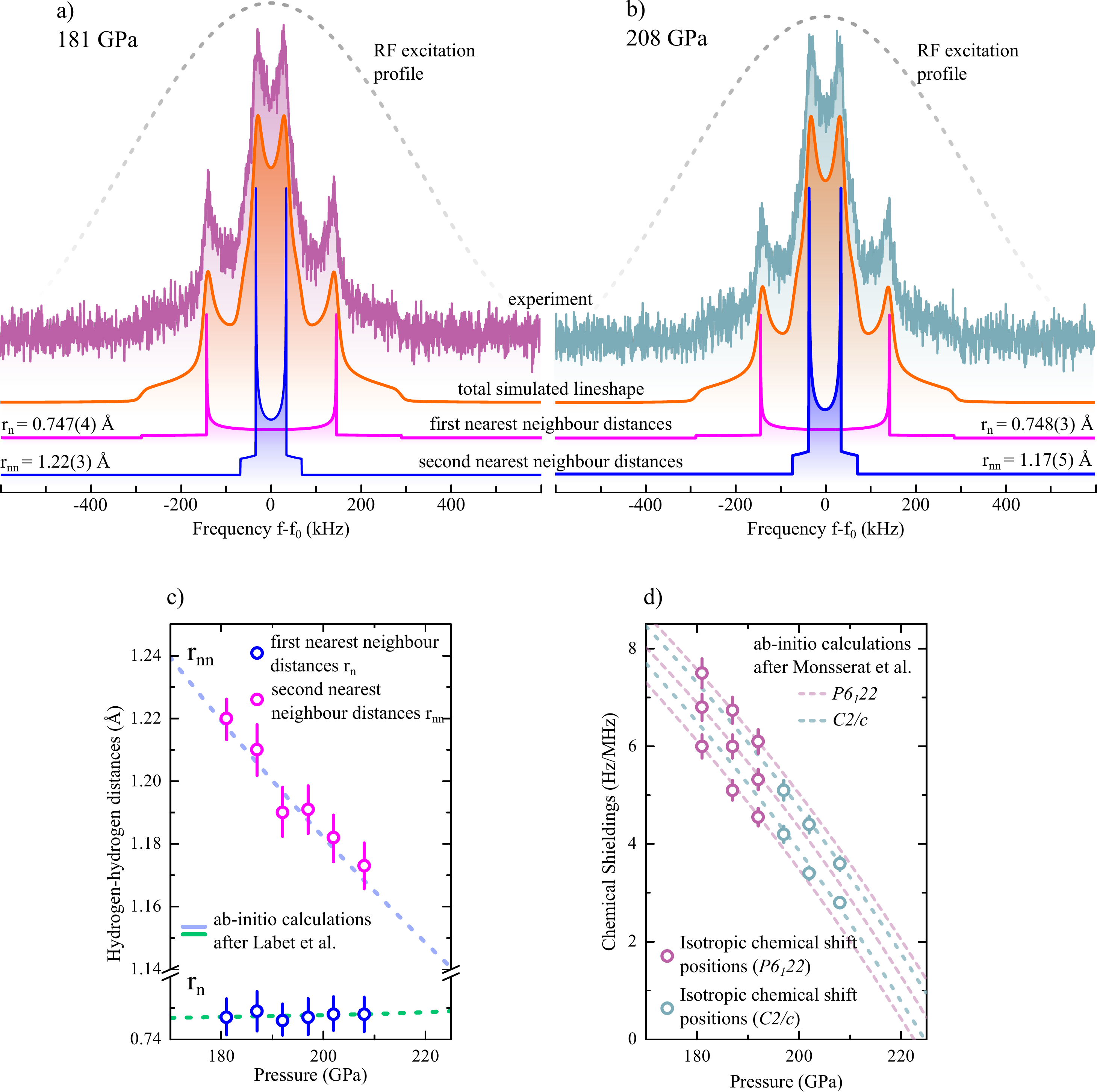}
\caption{\textbf{Data summary of hydrogen phase III} \textbf{a)} and \textbf{b)} $^1H$-NMR solid echo spectra at 181 GPa and 208 GPa. Recorded signals are a superposition of dipolar broadened Pake doublets \cite{Pake1948}, with spectral singularity splittings determined by first- and second nearest neighbour distances $r_n$ (pink) and $r_{nn}$ (blue), respectively. The resulting total lineshape (purple for 185 GPa and olive for 208 GPa) has been broadened by 10 kHz and was found to fit the experimental results well. For spectral simulation, we included the radio-frequency (RF) excitation profiles (grey) in order to accommodate for the detection of broad spectra. \textbf{c)} Comparison of $r_n$ and $r_{nn}$ with \textit{ab-initio} calculations from Labet et al. \cite{Labet2012}. Both $r_n$ and $r_{nn}$ do not show significant changes throughout the whole investigated pressure range and fit well with the theoretical predictions at those pressures. \textit{d)} Comparison of experimentally found isotropic chemical shielding values from the high-resolution Lee-Goldburg spectra with \textit{ab-initio} calculations from Monserrat et al.\cite{Monserrat2019}. The transition from $P6_1 22$ to $C2/c$ structures is evident by a sudden transition from triplet spectral features to a pronounced doublet structure. See text for more information. 
}
\label{fig1}
\end{figure*}
Recent methodological advancements in the field of high-pressure NMR using two-dimensional magnetic flux transformers\cite{Meier2017, Meier2018c} for directed excitation and sensitive detection of the faint nuclear spin response after radio-frequency excitation have allowed for a significant enhancement of accessible pressures into the mega-bar regime\cite{Meier2019a}, with a plethora of novel observed phenomena\cite{Meier2018ab, Meier2020a, Meier2022a}. 
\\
It was found that at pressures above about 70 GPa, the nuclear spin statistics of molecular hydrogen undergoes a distinct crossover from a $I=1$ (ortho-hydrogen) to a classical $I=1/2$ spin system as nuclear wave functions of inter- and intramolecular hydrogen atoms overlap significantly, ultimately leading to a collapse of the quantum-mechanical summation of intramolecular angular momenta, i.e. the disappearance of the nuclear spin isomers on NMR time scales \cite{Meier2020}. This phenomenon leads to distinct hydrogen spectra which are a superposition of two overlapping Pake doublets\cite{Pake1948}, see Figure \ref{fig1}, often with line-widths in excess of several hundred Hz/MHz or ppm, precluding the possibility of crystal structure determination due to lack of spectral resolution.
\\
\begin{figure*}[ht]
\centering    
\includegraphics[width=.95\textwidth]{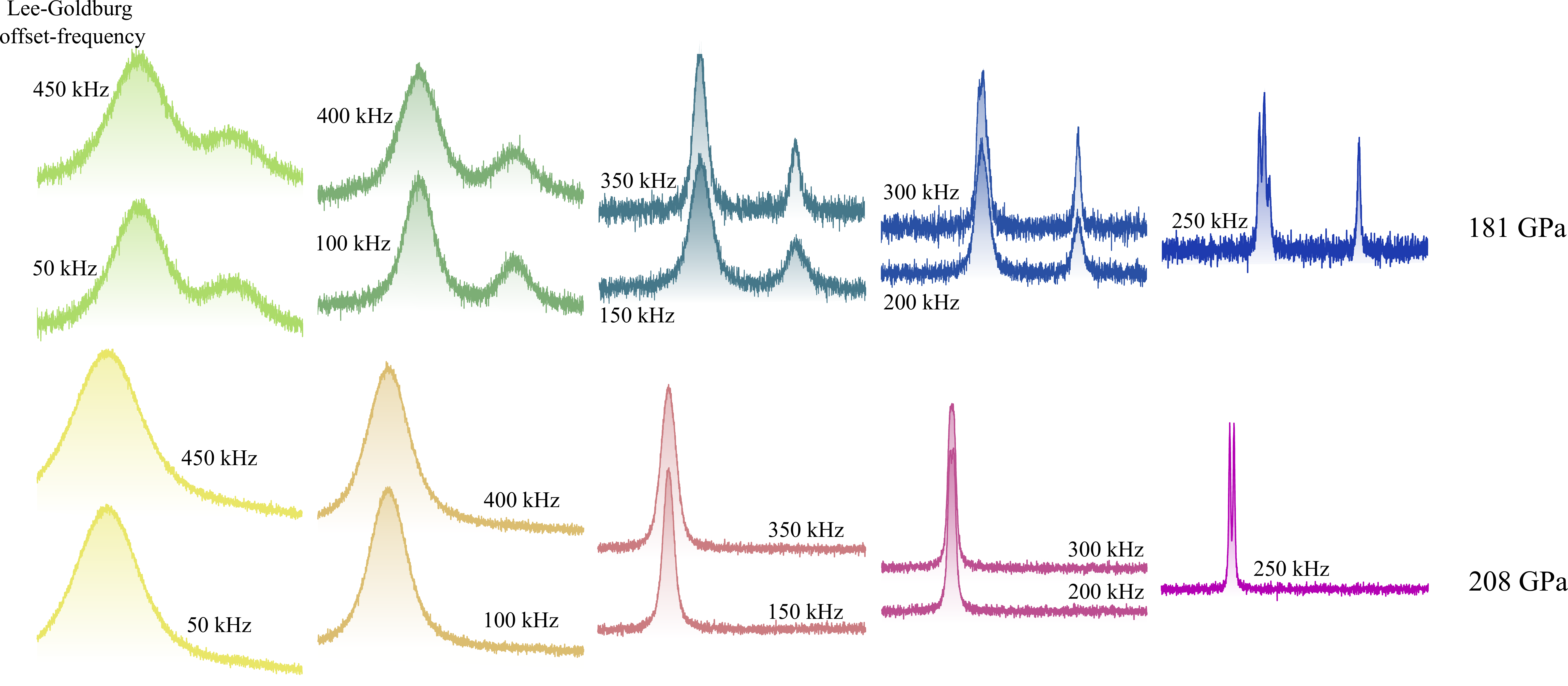}
\caption{\textbf{Lee-Goldburg NMR nutation spectra.} In order to acquire optimal spectral resolutions, several spectra at varying off-set frequencies have been recorded. Optimal conditions were found at 250 kHz, where individual spectral features at 181 and 208 GPa could be resolved. Spectral resolutions of $< 1~Hz/MHz$ could be achieved, allowing for the distinction of computed NMR spectra of the high pressure polymorphs of phase III. 
}
\label{fig2}
\end{figure*}
However, Meier et al. could demonstrate that high-resolution NMR is possible with the application of off-resonance RF pulses prior to excitation, forcing the precessing nuclear spin system to relax in the so called "Magic angle"\cite{Meier2019}, with achievable linewidths in the order of $10^{-1}$ to $10^1$ Hz/MHz \cite{Meier2021a}, leading to local magnetic field resolutions in the order of $10^0$ to $10^2$ $\mu T$ at common NMR field strengths.
\\
Here, we present the first combined high-pressure high-resolution NMR study of dense hydrogen phase III at room temperature from 181 to 208 GPa. Diamond anvil cells for hydrogen NMR experiments have been prepared similarly to previous work\cite{Meier2019, Meier2020}. 
\\
Figure \ref{fig1} shows a summary of $^1H$-NMR spin echo spectra. Direct excitation of NMR spectra leads to very broad hydrogen signals, with line widths of about 400 to 500 kHz, which are in accordance with previous high-pressure NMR studies up to 123 GPa\cite{Meier2020}. Lineshapes were simulated under the assumption of purely dipolar broadening as the leading spin interaction in this system. Additionally we included mild spectral distortion effects due to limited RF excitation of the outer lying spectral regions which are otherwise difficult to excite properly.  As can be seen, the resulting total calculated lineshapes fit the experimental spectra very well with extracted first ($r_n$) and second ($r_{nn}$) nearest neighbour distances in agreement with \textit{ab-initio} calculations from Labet et al.\cite{Labet2012}, see Figure \ref{fig1}c. Both $r_n$ and $r_{nn}$, and by extrapolation unit cell volumes,  were found to not change significantly over the investigated pressure range, in accordance with recent diffraction studies\cite{Ji2019}.  
\\
In order to quench the observed strong dipolar couplings of the hydrogen spectra in the pressure range from 181 to 208 GPa, Lee-Goldburg (LG) decoupling experiments have been employed. Figure \ref{fig2} depicts the spectral evolution under continuous decoupling for varying offset-frequencies $\Delta\omega_{off}$.
\\
Strikingly, even at values of $\Delta\omega_{off}$ far away from optimal conditions, i.e. at 50 and 450 kHz, the spectra at 181 GPa and 208 GPa show distinct spectral differences which were not observable in the spin echo spectra (Fig \ref{fig1}a and b). Further approach towards the LG optimum demonstrates the successive quenching of dipolar interactions in the nuclear spin Hamiltonian until at $\Delta\omega_{off}=~250~kHz$ two spectra with distinct spectral features emerge, with linewidths of about 240 and 280 Hz at 181 GPa and 208 GPa respectively. The achieved spectral resolution of about 0.5 Hz/MHz is in good accordance with other recent high-pressure high-resolution studies\cite{Meier2021a, Meier2022a}.   
\\
The spectrum at 181 GPa exhibits two distinct signals, a single sharp peak at 28 Hz/MHz and a triplet structure at chemical shieldings between 5 to 8 Hz/MHz, Figure \ref{fig3}a.
\\
\begin{figure*}[ht]
\centering    
\includegraphics[width=1\textwidth]{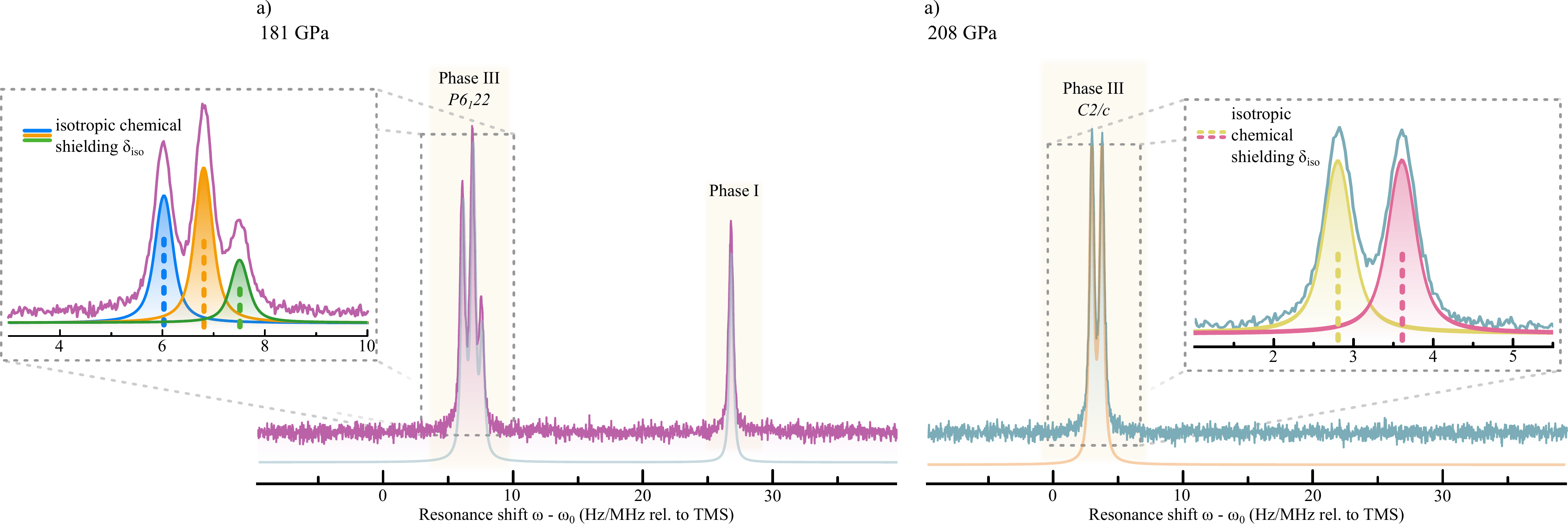}
\caption{\textbf{High-resolution spectra of hydrogen phase III.} \textbf{a)} $^1H$-LG-NMR spectrum at 181 GPa. Two distinct signals were found at about 28 Hz/MHz and a triplet at 10 Hz/MHz respectively. The latter singlet could be associated to hydrogen phase I by extrapolation of previous high-pressure NMR data\cite{Meier2020}. The triplet at 10 Hz/MHz was found to agree well with DFT calculations for a $P6_1 22$ space group according to \cite{Monserrat2019}. \textbf{b)} Hydrogen spectra above 197 GPa were found to exhibit a pronounced doublet structure originating in partial degeneracy for the four crystallographically non-equivalent sites in a monoclinic $C2/c$ structure, in excellent agreement with previous computational studies. Purple (181 GPa) and olive (208 GPa) are the resulting Fourier transform NMR spectra after Lee-Goldburg decoupling. Underlying solid lines are the respective spectral simulations. The insets show close-ups of NMR spectra associated to the hexagonal (181 GPa) and monoclinic (208 GPa) structures, respectively. The dotted lines are the isotropic chemical shielding values acquired during spectral simulation. Each spectrum was broadened by about 0.6 - 0.7 Hz/MHz. 
}
\label{fig3}
\end{figure*}
Extrapolating recorded chemical shielding values from our earlier studies of molecular hydrogen, see Figure 3c in \cite{Meier2020}, we conclude the sharp single peak at higher shieldings to be associated with phase I. This assumption is justified since molecular hydrogen phase I, in its simple hexagonal phase, exhibits only a single crystallographic position of the hydrogen quantum rotors, leading to a single sharp peak. Furthermore, it has been reported that within a pressure range of 170 - 180 GPa, phases I and III can coexist, gradually transforming phase I into phase III until about 190 GPa. The found intensity ratio between the singlet and the triplet has been found to be close to 3:1, in good accordance with high-pressure Raman studies\cite{Akahama2010EvidenceGPa}.
\\
The triplet at lower shieldings (see inset of Figure \ref{fig3}a) could be deconvoluted into three distinct Voigitian spectral lines of about 0.4 Hz/MHz linewidths, with isotropic chemical shield values of 6.0(2), 6.8(3) and 7.5(3) Hz/MHz, respectively. The found intensity ratio between each signal contributing to the triplet was found to be roughly 1.9 : 3.4: 1. If compared with \textit{ab-initio} calculations from Monserrat et al.\cite{Monserrat2019} we find that the experimentally observed spectrum is in very good agreement with the hexagonal $P6_1 22$ structural candidate of phase III. The doublet observed at 208 GPa was deconvoluted into two separate isotropic chemical shielding values of 2.8(1) and 3.6(1) Hz/MHz. Considering an intensity ratio close to unity, we conclude that the spectrum at 208 GPa is best described by the monoclinic structural candidate $C2/c$. 
\\
\begin{figure}[ht]
\centering    
\includegraphics[width=.6\columnwidth]{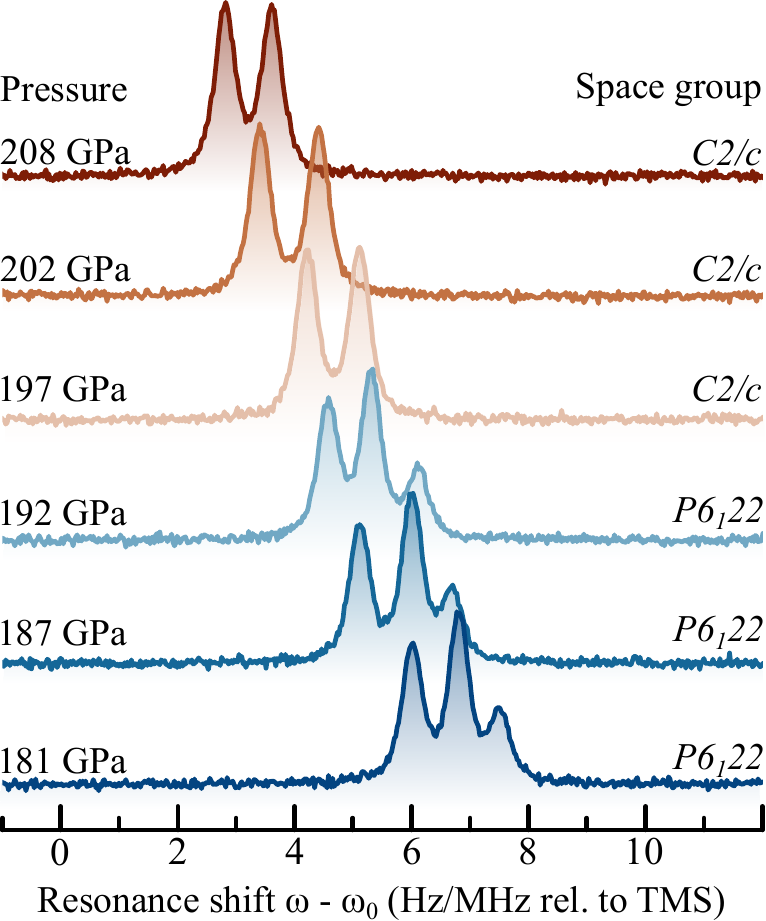}
\caption{\textbf{Summary of high resolution spectra of phase III.} From 181 to 192 GPa, a characteristic triplet structure of the $P6_1 22$ polymorph could be observed which transformed at 197 GPa into the observed monoclinic doublet NMR spectrum. The continuous up-field shift of the spectra is in good accordance with DFT calculations, see Figure \ref{fig1}.
}
\label{fig4}
\end{figure}
Figure \ref{fig4} shows the summary of all recorded $^1H$-NMR spectra from 181 GPa to 208 GPa, and Figure \ref{fig1}d) compares extracted isotropic chemical shielding values with the theoretical predictions of both the hexagonal and the monoclinic structural candidates of phase III. As can be seen, experimentally found high resolution values of isotropic shieldings fit very well with theoretical predictions of both $P6_1 22$ and $C2/c$ structural candidates, with a transition from hexagonal to monoclinic symmetry at about 195 GPa. 
\\
At this point it has to be mentioned that spectral predictions of both high pressure candidates overlap, as illustrated in Figure \ref{fig1}d. However, the appearance of line shapes which represent two distinctly different atomic arrangements can be considered unique, and the transition from one structure to the other is obvious. The up-field drifts (toward lower shielding values) in both $P6_1 22$ and $C2/c$ have been successfully predicted under increasing pressure as well. The overlap with experimental values was found to be reasonably good.
\\
It was shown\cite{Monserrat2019} that zero-point quantum renormalisation of the chemical shielding tensors will lead to a pronounced up-field shift of isotropic chemical shielding values. Thus, the good agreement between experimentally observed NMR spectra and \textit{ab-initio} DFT calculations indicate that zero-point motion plays a crucial role in the overall crystal structure prediction of high pressure phases of hydrogen. 
\\
Summarizing, the presented work illustrates firstly that the combination of \textit{ab-initio} DFT calculations and high-pressure high-resolution NMR is indeed capable of refining structures of low-Z materials like dense hydrogen. Secondly, the presented data experimentally confirms that hydrogen phase III undergoes a symmetry breaking from a hexagonal $P6_1 22$ to a monoclinic $C2/c$ structure.
\\
Owing to already described experimental challenges in the investigation of high-pressure hydrogen, this transition would be almost invisible using standard spectroscopic or diffraction methods. As has been shown earlier\cite{Monserrat2016}, Raman and infrared spectra of both structural candidates are very similar and diamond anvils used for ultra-high pressure generation often exhibit extensive luminescence\cite{Eremets2017}, prohibiting a conclusive distinction using optical or vibrational spectroscopy. 
\\
A similar situation occurs when attempting to conduct X-ray diffraction on phase III. As already mentioned, lack of core electron density leads to rather low interaction cross sections and thus very low signal amplitudes after a scattering event. This often leads to diffraction peaks barely visible above noise levels with regularly only the most prominent diffraction peaks visible\cite{Ji2019, Ji2020a}, which have almost identical positions for $P6_1 22$ and $C2/c$.
\\
Therefore the here presented novel approach for crystal structure determination for low-Z materials employing a synergy between NMR spectroscopy and \textit{ab-initio} DFT calculations was demonstrated to be fruitful in answering one of the most outstanding questions in the field of high-pressure hydrogen research, and will undoubtedly lead to more refined structural insights into the other high-pressure polymorphs of the most abundant element in the universe.
\\

\bibliographystyle{unsrtnat}
%

\section*{Acknowledgements}

We would like to acknowledge fruitful discussions with Ho-kwang Mao and Yang Ding. This work was supported by the National Science Foundation of China (42150101) and the National Key Research and Development Program of China (2022YFA1402301).  T. Meier acknowledges financial support from the Center for High Pressure Science and Technology Advanced Research as well as the Shanghai Key Laboratory \textit{MFree} and the Institute for Shanghai Advanced Research in Physical Sciences, Pudong, Shanghai.   

\section*{Author Contributions Statement}
Idea and Conceptualisation: T.M., M.Y., Y.Z.; Methodology:  T.M. and M.Y.; Investigation: M.Y., Y.Z, R.J., T.N., Y.F and T.M.; Visualization: M.Y., and T.M.; Formal analysis: M.Y., T.M.; Validation: M.Y.,Y.Z., R.J., T.M.; Funding acquisition: T.M.; Resources: T.M.; Project administration: T.M.; Writing—original draft: M.Y. and T.M.; Writing—review and editing: M.Y., Y.Z., R.J., T.N., Y.F, T.M.

\section*{Competing Interests Statement}

We declare no competing interest.

\section*{Data and code availability}

All raw data and python codes used to reach the here declared conclusions are available upon request.

\end{document}